\documentstyle[preprint,aps]{revtex}
\begin{document}
\draft
\title{Phase Transitions in Quantum Dots }

\author{H.-M. M\"uller and S. E. Koonin}
\address{W. K. Kellogg Radiation Laboratory, 106-38, California
Institute of Technology\\ Pasadena, California 91125 USA\\}
\date{\today}
\maketitle

\begin{abstract}
We perform Hartree-Fock calculations to show that quantum dots (i.e. 
two dimensional systems
of up to twenty interacting electrons in an external parabolic potential)
undergo
a gradual transition to a spin-polarized Wigner crystal
with increasing magnetic field strength.
The phase diagram and ground state energies  have been determined.
We tried to improve the ground state of the Wigner crystal
by introducing a Jastrow ansatz for the wavefunction
and performing a variational Monte Carlo calculation.
The existence of so called magic numbers
was also investigated. Finally, we also calculated
the heat capacity associated with the
rotational degree of freedom of deformed many-body states.
\end{abstract}

\narrowtext

\section{Introduction}
Quantum dots have been subject of recent intense experimental
and theoretical research. The interest in those nanostructures
arises not only from possible new technological applications, but also from the
desire to understand the fundamental physical problem of a few
($m \leq 300$)  interacting electrons in an
external potential and a strong magnetic field.

In a weak magnetic field
electrons form a rotationally symmetric state by occupying the
lowest Fock-Darwin
levels. With increasing magnetic field, the behaviour is determined by
three basic
mechanisms: the spin alignment of the electrons (eventually resulting in a
spin polarized system), the
Pauli principle (which causes the ground states of the system to occur with
certain ``magic numbers'' of the total angular momentum \cite{johns,maksym}),
and finally the Coulomb interaction.
This latter makes the electron droplet susceptible to
edge excitations and bulk instabilities, as the electron-electron interaction
favors a larger area \cite{MacDo,deCham}.

The scenario above assumes a unbroken rotational symmetry.
Questions concerning a high-field transition to so-called Wigner
molecules and crystals, in which electrons occupy fixed sites in a rotating frame and
are therefore localized, have been investigated
by Maksym \cite{maksym} and Bolton et. al.
\cite{bolton2}. Maksym considers a `large angular momentum limit' of systems
of up to
five electrons and describes only excited states of integer angular momentum.
He speculates on the existence of ground state Wigner molecules
in the large field limit.
Bolton et. al. simulated
up to forty \emph{classical} interacting point charges in an external
parabolic potential.

In this article we consider the ground state properties
of up to twenty electrons in the limit of a 
strong magnetic field. We treat the full quantal
problem by solving the Hartree-Fock equations for the two-dimensional
electron gas.
After a brief description of the well known phases of the rotationally
symmetric case as they appear in the Hartree-Fock approximation, we present
the gradual transition towards a Wigner molecule and crystal;
the various spatial configurations are shown and the shell structure is 
compared to that of the classical calculation \cite{bolton2}.
We further describe a variational Monte Carlo calculation which we performed
to improve the wave function by a Jastrow ansatz.
Finally, we investigate the rotational spectra associated with the breaking of
the continuous rotational symmetry; the heat capacity associated with this new
rotational degree of freedom is calculated.

\section{Theory}
We consider $m$ electrons of effective mass $m^\ast$
in a plane $(x,y)$ confined by an external parabolic potential,
$ V(r) = \frac{1}{2} m^\ast \omega_0^2 r^2$, and subject to a strong
magnetic field $\vec{B} = B_0 \vec{e}_z$. The Hamiltonian for such a system
is
\begin{equation}
\hat{H} = \sum_{i=1}^{m} \frac{1}{2m^\ast} \vec{\Pi}_i^2 + \sum_{i=1}^{m}
\frac{1}{2} m^\ast \omega_0^2 (x_i^2 + y_i^2)
+ \sum_{i=1}^m  \frac{g^\ast \mu_B \vec{B} \cdot \vec{S_i}}{\hbar}
+ \sum_{i < j} \frac{e^2}{\epsilon | \vec{r_i} - \vec{r_j} | },
\end{equation}
where $ \vec{\Pi}_i = \frac{\hbar}{i} \vec{\nabla_i} + \frac{e}{c}
\vec{A}(\vec{r_i})$ is the kinetic momentum of the $i$th electron in the
vector potential $\vec{A}(\vec{r_i}) = \frac{B_0}{2}(-y_i,x_i,0)$.
We include the spin degree of freedom of the electrons, $\vec{S_i}$
(which implies the addition of the Zeeman energy
with $g^\ast \approx 0.54$ as the effective $g$ factor for the material GaAs),
but neglect any spin-orbit interaction. (An order-of-magnitude
estimate for the magnetic field strength induced by the circular motion
of an electron with an angular momentum of ${\rm \hbar}$ at
a distance of $l_0 \sim 1.0 \times 10^{-6} \;{\rm cm}$ is only
$\frac{e \hbar}{m^\ast c l_0^3} \approx 2.8 \times 10^{-5} \; {\rm T}$.)
Defining the frequencies $\omega_c = \frac{eB_0}{m^\ast c}$ and
$\omega(B_0) = \sqrt{\omega_0^2 + \frac{1}{4} \omega_c^2}$, we rewrite
the coordinates as dimensionless complex variables,
$z_i = \frac{1}{l_0} \frac{x_i - iy_i}{\sqrt{2}}$, where
$l_0 = \sqrt{\frac{\hbar}{m^\ast \omega(B_0)}}$  is the
magnetic length, to obtain
\begin{eqnarray}
\hat{H}  & = &\sum_{i=1}^{m} - \frac{\hbar^2}{m^\ast} \left( \frac{1}{l_0^2}
\partial_{z_i} \partial_{\bar{z_i}} \right) + m^\ast \omega^2(B_0) l_0^2
|z_i|^2
+ \frac{\hbar}{2}\omega_c (\bar{z_i}\partial_{\bar{z_i}}
- z_i\partial_{z_i}) \nonumber \\ 
& + & \sum_{i=1}^m  \frac{g^\ast \mu_B \vec{B} \cdot \vec{S_i}}{\hbar}
+ \sum_{i < j}
\frac{e^2}{\epsilon \sqrt{2} l_0 | z_i - z_j | } 
= \sum_{i=1}^{m} \hat{H_0}(z_i,\vec{S_i})
+ \frac{e^2}{\epsilon \sqrt{2} l_0}
\sum_{i < j} \frac{1}{ | z_i - z_j | }.
\end{eqnarray}

$\hat{H_0}$ is the single-particle Hamiltonian whose eigenfunctions will form
the basis states for the Hartree-Fock calculation. For the spatial part
of  $\hat{H_0}$ we define
\begin{equation}
a^\dagger = \frac{1}{\sqrt{2}} (\bar{z} - \partial_z), \; b^\dagger =
\frac{1}{\sqrt{2}} (z - \partial_{\bar{z}}), \;
a = \frac{1}{\sqrt{2}} (z + \partial_{\bar{z}}), \; b = 
\frac{1}{\sqrt{2}}
(\bar{z} + \partial_z)
\end{equation}
with $[a,a^\dagger] = [b,b^\dagger] = 1$ and write
\begin{equation}
\hat{H_0}(z) = \hbar \omega(B_0) (a^\dagger a + b^\dagger b + 1) - \frac{1}{2}
\hbar\omega_c(b^\dagger b - a^\dagger a) = \hbar\omega(B_0)
(2 a^\dagger a + {\cal L} + 1)   - \frac{1}{2} \hbar \omega_c {\cal L},
\end{equation}
where ${\cal L} := b^\dagger b - a^\dagger a$ is the angular momentum of
the particle.
The eigenvalues of the single particle Schr\"odinger equation
$\hat{H_0} \Psi_{n k} = \epsilon_{n k} \Psi_{n k}$,
$\epsilon_{n k} = \hbar\omega(B_0) \{ 2n + |k| + 1 \} - \frac{1}{2}
\hbar \omega_c k$,
indicate that for strong magnetic fields, e.g. $B_0 \approx 10 \;{\rm T}$,
one has $\hbar \omega_c \approx 17 \;{\rm meV}$ and
$\hbar\omega(B_0) \approx 9 \;{\rm meV}$,
and all particles occupy the Fock-Darwin states with $n = 0$:
\begin{displaymath}
\Delta \epsilon = \epsilon_{n=0 L} - \epsilon_{n=0 L-1} \approx 0.5 \;
{\rm meV}
\ll  \epsilon_{n=0 L} - \epsilon_{n=1 L} \approx 18 \;{\rm meV}.
\end{displaymath}
We therefore restrict our calculation to the $n = 0$
level, which
resembles the lowest Landau level if the external potential had been switched
off. The eigenfunctions are $\Psi_k \sim z^k \exp (- |z|^2)$ and are
identical to the usual
form $\phi_k \sim e^{-ik\varphi} r^k L_0^{|k|}(r)\exp( -\frac{r^2}{2l_0^2})$,
$L_0^{|k|}(r)$ being the Laguerre polynomial of degree zero.

In the Hartree-Fock calculation, we minimize the Hartree-Fock energy
\begin{eqnarray}
\label{ehf}
E^{HF} & = & \langle\Phi\vert \hat{H} \vert \Phi \rangle
 = \sum_{l_1 l_2} t_{l_1 l_2} \langle\Phi\vert c_{l_1}^\dagger c_{l_2}
\vert \Phi \rangle + \frac{1}{4} \sum_{l_1 l_2 l_3 l_4}
\bar{v}_{l_1 l_2 l_3 l_4}
\langle\Phi\vert c_{l_1}^\dagger c_{l_2}^\dagger c_{l_4} c_{l_3}
\vert \Phi \rangle \\
& = &  \sum_{l_1 l_2} t_{l_1 l_2} \rho_{l_2 l_1} + \frac{1}{2} 
\sum_{l_1 l_2 l_3 l_4} \rho_{l_3 l_1} \bar{v}_{l_1 l_2 l_3 l_4} \rho_{l_4 l_2}
\nonumber
\end{eqnarray}
with $ \rho_{l_1 l_2} = \langle\Phi\vert c_{l_2}^\dagger c_{l_1}
\vert \Phi \rangle$ being the density matrix 
and $\vert \Phi \rangle$ being a Slater determinant.
$c_{l}^\dagger$ creates a fermion in the state $\Psi_l$,
while its Hermitian conjugate $c_{l}$ destroys it.
The indices $l_i=(k_i,s_i)$ run over all orbital states $k$,
as well as the spin degree of freedom $s = \{+\frac{1}{2},-\frac{1}{2} \}$.
$t_{l_1 l_2} = \langle \Psi_{l_1} \vert \hat{H_0}  \vert\Psi_{l_2}\rangle
= \epsilon_{k_1} \delta_{k_1 k_2} + (-1)^{s_1 + \frac{1}{2}}
\frac{g^\ast \mu_B B_0}{2} \delta_{s_1 s_2}$
is the single particle matrix element of the Hamiltonian and
$\bar{v}$ the antisymmetrized Coulomb matrix element,
\begin{equation}
\bar{v}_{abcd} = \frac{e^2}{\epsilon \sqrt{2} l_0} \left\{ \begin{array}{ll}
\langle ab \vert \frac{1}{| z_i - z_j |} \vert cd \rangle
-\langle ab \vert \frac{1}{| z_i - z_j |} \vert dc \rangle 
& \text{if $s_c = s_d$,}\\
\langle ab \vert \frac{1}{| z_i - z_j |} \vert cd \rangle
& \text{if $s_c \neq s_d$.} \end{array}
\right.
\end{equation}

To minimize (\ref{ehf}), we vary with respect to
$\rho$, $ \delta E^{HF} / \delta \rho = 0$, with the contraints that we stay
within the set of Slater determinants ($\rho^2 = \rho$) and conserve the 
number of particles (${\rm tr} \rho = m$)
resulting in the matrix diagonalization problem
\begin{equation}
\label{diagh}
\sum_j h_{ij} D_{jk} = \sum_{j} \left( t_{ij} + \Gamma_{ij} \right) D_{jk}
= \sum_j \left( t_{ij} + \sum_{ll^\prime} \bar{v}_{il^\prime jl}
\rho_{ll^\prime} \right) D_{jk} = \varepsilon_k D_{ik},
\end{equation}
where $\Gamma$ is the so-called mean field. The eigenvectors $\vec{D}_k$ of
$h$ represent the new single-particle states $\{k\}$, that are to be
occupied according to the energies $\varepsilon_k$.
Equation (\ref{diagh}) has to be solved self consistently, since
$\rho_{ll^\prime} = \sum_{i=1}^{m} D_{li} D^*_{l^\prime i}$.

We use the Fock-Darwin representation in our calculation and take into
account up to $200$ single-particle states (including spin).
We tested our code by comparison with the results of
Pfannkuche et. al. \cite{pfanne} and Bolton \cite{bolton}.
Although Pfannkuche et al. describe quantum dot helium in a
model space that is different from ours
(they included $n \neq 0$ states in their calculation), our
ground state energies of total angular momentum $J = 1$ 
for $0 \; {\rm T} \leq B_0 \leq 5 \; {\rm T}$
coincide with their Hartree-Fock calculation within less then $2\%$, and the
$J = 0$ ground state energies agree with less than $5\%$. As one can see in
table I of reference \cite{pfanne}, the $n \neq 0$ coefficients in their
$J = 0$ ground state are larger than in their $J = 1$ ground state,
so that the $n \neq 0$ space is more significant for those 
magnetic field strengths.
Similar results are obtained if we compare our results to the fixed node Monte
Carlo calculation of reference \cite{bolton}. In the spin polarized case
our ground state energies agree within a few percent, while we
overestimate the energy of the depolarized system by up to $15\%$.
This is due to the larger correlation energies (ignored in a Hartree-Fock
calculation) when
two electrons can occupy the same orbital. Since the questions
addressed in this article concern the spin polarized regime,
this deviation from the results of \cite{bolton} is of little concern.

The Hartree-Fock approximation is known to conserve symmetries present
in the initial trial wavefunction. To generate deformed solutions,
we started with a quite arbitrary, but not rotationally invariant,
initial Slater determinant, which produces a deformed initial mean field.
Self consistent iteration of the Hartree-Fock scheme guarantees amplification
of solutions with the symmetry of the Wigner molecule. Of course, the
same converged solution must be reached for several different initial states
to give confidence that it is the true minimum.

\section{Numerical Results}
We have used the material constants of GaAs (i.e., $m^\ast = 0.067 \;m_e$ and 
$\epsilon = 12.9$, as well as an external potential strength of
$\hbar \omega_0 = 3 \; {\rm meV}$) for our calculation. 
To observe the expected phase transition, we first consider a system of
$m=10$ electrons. Fig \ref{fig1} shows the ground state energy as a function
of the magnetic field strength $B_0$ and for comparison the lowest energy
of the rotationally symmetric system. The Wigner molecule becomes the ground
state for $B_0 \gtrsim 5.2 \; {\rm T}$, while at smaller strengths the
rotationally symmetric state is favored. The system undergoes
spin polarization from $B_0 = 0 \; {\rm T}$ to $1.5 \; {\rm T}$, where
the spin polarized so-called maximum density droplet \cite{MacDo} prevails.
At $B_0 = 4.5 \; {\rm T}$ bulk instabilities result in unoccupied
inner Fock-Darwin states. The transition to a Wigner molecule, and later to a
crystal, happens very gradually. (We refer to the case where the
probability density is deformed, but still very smeared out as a
``molecule'', while a ``crystal'' signifies well localized and
distinguishable electrons, as illustrated in Fig \ref{fig2}.)
The molecule at $B_0 = 6 \; {\rm T}$ is lower in energy by only
$0.2 \% \; (0.542 \; {\rm meV})$
relative to the rotationally symmetric solution, while the crystal at
$B_0 = 10 \; {\rm T}$ gains about $\sim 3 \; {\rm meV}$,
which is of order of the strength of the confining potential. Note that the
deformed ground states are not eigenstates of the total angular momentum
operator $J = \sum_i {\cal L}^{(i)}$.

The rotationally symmetric case suffers further complication with increasing
magnetic field strength: While at first ($B_0 \approx 6 \; {\rm T}$) the hole
in the bulk widens (the $l=1,2,3$ Fock-Darwin levels empty) and
later at $B_0 = 6.75 \; {\rm T}$ a fourth state empties, the solution
transforms into two separate rings at $B_0 \geq 9 \; {\rm T}$.

For further insight into the various transitions, the radial
(angle-averaged) particle distribution for various
magnetic fields is shown in Fig \ref{fig3} for $m=20$ electrons.
The crystalline state has one electron in the
center of the dot, seven in a middle ring and 12 electrons in the outmost
region. Correspondingly, the $B_0 = 20 \; {\rm T}$ curve shows three maxima.
For $B_0 = 6 \; {\rm T}$ the center electron and the seven in the middle
ring have almost uniformly merged to a flat distribution which extends to
$z \approx 2$, and the outer ring can now be found at $z \approx 3$.
For $B_0 = 3 - 4 \; {\rm T}$ we find again the so-called maximum density
droplet: the electrons occupy the first twenty Fock-Darwin levels, since
they are polarized.
Further lowering of $B_0$ results in a depolarization, allowing
further accumulation of electrons near the origin. Since we only take into
account $n=0$ states, we cannot claim to represent the physical situation
for the smaller field strength, although we do reproduce the energies in this
regime quite well, as noted above.

In Fig \ref{fig4} we plot the separation energy
$\Delta(m) = E_{m+1} - E_{m}$ and the differences in the
separation energy, $\Delta_2 (m) = \Delta(m+1) - \Delta(m)$, as functions of
the particle number $m$ in the crystal regime, $B_0 = 20 \; {\rm T}$.
There is a large drop in $\Delta_2$ of $\sim 0.5 \; {\rm meV}$ 
whenever charge can be put to the outer region
of the dot (see, e. g., $m = 4$ and $m = 8$), in accord with
charge being distributed over a larger area, thereby
reducing the Coulomb energy. In the case where one charge is placed in the
center and two rings outside ($m = 14$), the gain in energy is reduced
by the fact that more particles outside feel a stronger external potential.
The tendency here is that the Coulomb energy plays a less and less important
role, weakening the slope in the separation energy, combined with the fact that
one can pack more particle in the outer region.

For comparison, we also show in Fig \ref{fig4} $\Delta(m)$ and $\Delta_2 (m)$
for the lowest rotationally symmetric state. No clear tendency
in the behaviour of $\Delta_2 (m)$ is evident.
The system is frustrated by the particles having to occupy Fock-Darwin
levels.

In Table \ref{tab1} we show the spatial configurations of the system in the
Wigner-like structure (obtained by enumerating the number of electrons
occupying the corresponding rings) and give the ground state energies.
We generally confirm the configurations of the classical calculation of
reference \cite{bolton2} as well as the exceptional behaviour of the
$m=6,10,12$ and $17$ clusters, although there is no peak in 
$\Delta_2$ for $m=14$ (the peaks in Fig \ref{fig4} correspond to the
cusps of Fig 5 of reference \cite{bolton2}), since ten
electrons are moved outside for the $m=16$ configuration.

The Hartree Fock calculation is based on a theory of independent particles
moving in an average potential. We improved the wavefunctions for the Wigner
regime to a many body wavefunction by introducing a Jastrow type function
\begin{equation}
\vert\Psi \rangle = \left( {\cal S} \prod_{i<j} f(z_i - z_j) \right) \Phi_{HF}
\end{equation}
where ${\cal S}$ is the symmetrizer and $\Phi_{HF}$ the Hartree Fock solution
to the problem.
To guarantee a convenient symmetrized form of the product of these
function, we made the ansatz
\begin{equation}
f(z_i - z_j) = \vert z_i - z_j \vert^k
\end{equation}
for the pair correlation function $f(z_i - z_j)$ 
with $k$ as a variational parameter.
We performed a variational Monte Carlo calculation \cite{carlson} to evaluate
the energy
\begin{equation}
E \left[ k\right] = \frac{ \langle \Psi \vert H \vert \Psi \rangle }
{ \langle \Psi \vert \Psi \rangle}.
\end{equation}
As it turned out, the Jastrow type wavefunction did not significantly improve
the Hartree-Fock energy.
In the case of 10 electrons and $B = 20 \; {\rm T}$,
the energy could only be improved by
$0.1 \% \; (0.4 \;{\rm meV})$ at $k = 0.1$.
For $0.1 < k< 1$ the energy is slowly increasing,
while for $k>1$ highly excited states are simulated as more holes
are introduced into the wavefunction. Obviously,
the Hartree-Fock solution already describes the Wigner state accurately.

In Fig \ref{fig4a} we plot the phase diagram with respect
to number of particles and the ratio $\frac{\omega_c}{\omega_0}$.
We omit the regime of bulk
instabilities, since it is of minor importance. The phase boundary of
the spin polarized regime and the partially unpolarized regime suffers again
from the Hartree-Fock approximation, as it bends down with decreasing number
of electrons. The boundary of the molecular regime is defined by how much the
continuous rotational symmetry is broken: the fractional
uncertainty in the total angular momentum is $f=\frac{\Delta J}
{\langle \hat{J} \rangle}
= \frac{\sqrt{ \langle \hat{J}^2 \rangle - \langle \hat{J} \rangle^2}}
{\langle \hat{J} \rangle}$, and we define a molecule by $f > 1\%$.
The boundary is fairly constant for $m > 6$, but,
since the transition is gradual, it has some uncertainty.
For less than eight particles, we find a small drop in the boundary,
due either to some non-obvious physical effect
or to the approximation we use.

In our Hartree-Fock solutions of ten or more electrons and
$B_0 = 20 \; {\rm T}$, the relative uncertainty in total angular momentum, $f$,
is of order of $10\%$. As in atomic nuclei,
these deformed solutions give rise to rotational spectra, which do not
appear in the case of the unbroken symmetry. We have estimated the spectrum
of rotational excitations by projecting the Hartree-Fock Slater determinant
onto eigenfunctions of good angular momentum $I$ \cite{schuck}.
The projector has the form
\begin{equation}
\hat{P}^I = \frac{1}{2\pi} \int_0^{2\pi} e^{i\alpha (\hat{J} - I )}
{\rm d} \alpha
\end{equation}
and the energies which result from taking the mean value of $\hat{H}$ with the
projected wavefunctions are given by
\begin{equation}
\label{eproj}
E_{proj}^I = \frac{ \langle\Phi\vert \hat{H} \hat{P}^I \vert \Phi \rangle}
{\langle\Phi\vert \hat{P}^I \vert \Phi \rangle} = 
\frac{\int {\rm d} \alpha h(\alpha) e^{-iI \alpha}}
{\int {\rm d} \alpha n(\alpha) e^{-iI \alpha}},
\end{equation}
defining the quantities
$h(\alpha) =  \langle\Phi\vert \hat{H} e^{i \alpha \hat{J}} \vert \Phi \rangle$
and $n(\alpha) =  \langle\Phi\vert e^{i \alpha \hat{J}} \vert \Phi \rangle$.
Since the standard deviation in $\hat{J}$ is only of few percent, one can
calculate these matrix elements approximately by writing $h(a)$
in the expansion
\begin{equation}
\label{expansion}
h(\alpha) = \sum_{n=0}^K h_n \left(
- \langle \hat{J} \rangle + \frac{1}{i} \frac{\partial}{\partial \alpha }
\right)^n n(\alpha).
\end{equation}
One justifies this ansatz with the fact that it represents a Taylor
expansion of the Fourier transformed function $ h(\alpha) /  n(\alpha)$,
and, assuming that both quantities are sharply peaked at $\alpha = 0$,
this quotient is smooth and can be approximated by a few terms of equation
(\ref{expansion}). By operating $(\langle - \hat{J} \rangle + \frac{1}{i}
\frac{\partial}{\partial \alpha})$ on equation (\ref{expansion})
and setting $\alpha = 0$,
one gets an inhomogeneous system of equations for the unknown $h_0 ... h_K$:
\begin{equation}
\langle H ( \Delta \hat{J})^m \rangle = \sum_{n=0}^K h_n \langle
( \Delta \hat{J} )^{m+n} \rangle.
\end{equation}
Equation (\ref{eproj}) can then be expressed as
\begin{equation}
\label{eproj2}
E_{proj}^I = \sum_{n=0}^K h_n (I - \langle \hat{J} \rangle )^n.
\end{equation}
\\
We restrict ourselves to $K=2$, since higher terms involve the calculation
of $k$-body operators with $k>4$. For this case, we have
\begin{eqnarray}
h_2 & = & \frac{
\langle \Delta \hat{J}^2 \rangle \langle \hat{H} \Delta \hat{J}^2 \rangle 
- \langle \Delta \hat{J}^3 \rangle \langle \hat{H} \Delta \hat{J} \rangle 
- \langle \Delta \hat{J}^2 \rangle^2  \langle \hat{H}  \rangle}
{\langle \Delta \hat{J}^2 \rangle \langle \Delta \hat{J}^4 \rangle
- \langle \Delta \hat{J}^3 \rangle^2
- \langle \Delta \hat{J}^2 \rangle^3}
\nonumber \\
h_1 & = &  \frac{\langle \hat{H} \Delta \hat{J} \rangle}
{\langle \Delta \hat{J}^2 \rangle} - h_2
\frac{\langle \Delta \hat{J}^3 \rangle}{\langle \Delta \hat{J}^2 \rangle}
\nonumber \\
h_0 & = & \langle \hat{H} \rangle - h_2 \langle \Delta \hat{J}^2 \rangle.
\end{eqnarray}

Fig \ref{fig5} shows the rotational spectra for $m=10$ and $m=20$ electrons
and $B = 20 \; {\rm T}$ as a function of the quantum number $I$,
where we have substracted the shifted ground state energy, which
is obtained from the Hartree-Fock energy $\langle \hat{H} \rangle$ by substracting
the spurious rotational energy $h_2 \langle \Delta \hat{J}^2 \rangle$,
which is only of order $0.25 \; {\rm meV}$
in both cases. The moments of inertia associated with these states are
$J_Y = \frac{1}{2 h_2} = 5.2 \times 10^5 \; {\rm \hbar^2 / eV}$ for $m=10$ and
$J_Y = 1.9 \times 10^6 \; {\rm \hbar^2 / eV}$ for $m=20$.

In order to excite a molecule with circular polarized
radiation, one has to produce photons of minimal energy of
$\Delta E^{(10)}(I=224) = E_{proj}^I(I=224) - (\langle \hat{H} \rangle 
- h_2 \langle \Delta \hat{J}^2 \rangle) = 1.12 \cdot 10^{-7} \; {\rm eV}$
for the ten electron molecule and
$\Delta E^{(20)}(I=790) = 3.2 \cdot 10^{-8} \; {\rm eV}$  
for twenty electrons, which are the energy differences between ground and
first excited state.
These energies correspond to radio frequencies of $\nu^{(10)} = 27.06 \;
{\rm MHz}$
and $\nu^{(20)} = 7.73 \; {\rm MHz}$.
Note that the corresponding wavelengths are in the transparent region for GaAs.
Therefore the measurement of transmission coefficients of circular polarized
radiation should give experimental evidence of Wigner molecules.
The level spacing, $\Delta E \approx \frac{\partial E_{proj}^I}{\partial I}
\Delta I$, of the excited states then increases with higher states, resulting
in excitations in the microwave region.
The heat capacity connected with this rotational degree of freedom,
\begin{equation}
c = \frac{\partial <U>}{\partial T} =  \frac{\partial}{\partial T}
\frac{1}{Z}
\left(
\sum_I E_{proj}^I \exp \left( - \frac{E_{proj}^I}{k_B T} \right)
\right),
\end{equation}
where $Z =  1
+ \sum_I \exp \left(-E_{proj}^I / k_B T \right)$ is the partition
function and $k_B$ Boltzmann's constant, should therefore reach its classical
value of $\frac{1}{2} k_B$ even for temperature as low as $1 \; {\rm K}$.
Fig \ref{fig6} shows the well known Schottky anomaly of the heat capacity,
typical for a system where only two states are of importance, at low
temperatures of $\sim 1 \; {\rm mK}$. As expected, it approaches
$\frac{1}{2} k_B$ for high temperatures.
For the indicated temperature regime the heat capacity has
converged within our model space, which consists of 400 rotational states
and shows the expected typical behaviour of a quantum mechanical
rotor in a heatbath.

The energy levels of the vibrational modes of a single electron in the crystal
can be estimated in a simplified one dimensional model. Concerned only with
the radial degree of freedom, an outer electron (in the case of ten electrons)
interacts with the external potential and the Coulomb potential of the two
inner electrons, which we regard positioned at the center:
\begin{equation}
V(r) = \frac{1}{2} m^\ast \omega_0^2 r^2 + \frac{2e^2}{\epsilon r}.
\end{equation}
Expanding the potential around the equilibrium position $r_0$ of the outer
electron to second order, we obtain
\begin{equation}
V(r) = \frac{1}{2} m^\ast \left\{ \omega_0^2 
+ \frac{8e^2}{\epsilon m^\ast r_0^3} \right\} \left(r-r_0\right)^2
+ \left\{ m^\ast \omega_0^2 r_0 - \frac{2e^2}{\epsilon r_0^2} \right\}
\left(r-r_0\right) +\frac{1}{2} m^\ast \omega_0^2 r_0^2
+ \frac{2e^2}{\epsilon r_0}.
\end{equation}
The electron is confined by the parabolic part of this expansion with
an corrected strength $\omega^\prime = \sqrt{ \omega_0^2 +
\frac{8e^2}{\epsilon m^\ast r_0^3}}$. Setting $r_0 \approx 2 \times 10^{-6}
\; {\rm cm}$, the energy levels for the vibrational modes of the electron
are separated by
$\Delta E_{vib} = \hbar \omega(B_0) = \sqrt{ (\hbar \omega^\prime)^2
+ \frac{1}{4} (\hbar\omega_c)^2} \approx 21 \; {\rm meV}$,
much larger than the separation in the rotational energy levels
($\Delta E_{rot} \approx 10^{-4} \; {\rm meV}$
around $I = 225 \; {\rm \hbar}$).
Vibrational modes therefore contribute only marginally to the heat capacity
and can be easily suppressed by proper excitation of the rotational modes
only.

In summary, we have shown in a full quantum mechanical treatment that there
exist regimes where Wigner molecules and crystals are the ground states of 
quantum dots. We have also described rotational spectra of quantum dots,
which arise from the existence of deformed Hartree-Fock solutions. This
broken symmetry could make it possible to detect Wigner molecules
experimentally by exciting the rotational excited states of the system.
Open questions remain: How do the two-body correlations neglected 
in the Hartree-Fock approximation influence systems of few
electrons ($m < 5$) and what kind of state is formed in the case of
many particles ($m>40$)?

\acknowledgements
This work was supported in part by the National Science Foundation,  
Grants No. PHY94-12818 and PHY94-20470.
We thank Karlheinz Langanke and Ming Chung Chu for helpful discussions.

\begin{table}[ht]
\begin{center}
\begin{math}
\begin{array}[t]{|c|c|c|}
\hline
\rm number \; of & \rm energy & \rm ring \; occupations\\
\rm  electrons &\rm \left[ meV \right]  & \rm inner-middle-outer\\
\hline
1  & 17.247   & 1-0-0 \\
2  & 40.085   & 2-0-0 \\
3  & 66.439   & 3-0-0 \\
4  & 96.463   & 4-0-0 \\
5  & 129.986  & 5-0-0 \\
6  & 166.346  & 1-5-0 \\
7  & 205.448  & 1-6-0 \\
8  & 247.636  & 1-7-0 \\
9  & 292.621  & 2-7-0 \\
10 & 339.934  & 2-8-0 \\
11 & 389.489  & 3-8-0 \\
12 & 441.634  & 3-9-0 \\
13 & 496.008  & 4-9-0 \\
14 & 552.825  & 4-10-0 \\
15 & 611.879  & 5-10-0 \\
16 & 673.004  & 1-5-10 \\
17 & 736.135  & 1-5-11 \\
18 & 801.162  & 1-6-11 \\
19 & 868.558  & 1-6-12 \\
20 & 937.973  & 1-7-12 \\
\hline
\end{array}
\end{math}
\caption{\label{tab1}Ground state energies
and spatial distributions of Wigner crystal in quantum dots
for up to twenty electrons at $B=20 \; {\rm T}$.}
\end{center}
\end{table}

\begin{figure}
\caption{Ground state energy of the Wigner molecule (solid line) and
lowest rotationally symmetric state (dashed line) for ten electrons.}
\label{fig1}
\end{figure}

\begin{figure}
\caption{The electron density distribution for $m=10$ electrons. The left
panel shows the solution at $B_0= 6 \; {\rm T}$, the right at
$B_0= 10 \; {\rm T}$.}
\label{fig2}
\end{figure}

\begin{figure}
\caption{Radial electron density of $m=20$ electrons for different
values of $B_0$. The solid curves represent Wigner crystals, while the
dashed curves show the slow transition to a maximum density droplet,
which is drawn with a dashed-dotted line.
Depolarization sets in for the cases of the dotted curves.}
\label{fig3}
\end{figure}

\begin{figure}
\caption{Separation energy $\Delta(m)$ and the differences in the separation
energy $\Delta_2(m)$ for Wigner molecules (upper two diagrams) and for the
lowest available rotationally symmetric states (lower two diagrams).}
\label{fig4}
\end{figure}

\begin{figure}
\caption{Phase diagram for quantum dots, plotting $\frac{\omega_c}{\omega_0}$
versus the number of electrons $(m)$. The lines crudely trace the
boundaries to guide the eye.}
\label{fig4a}
\end{figure}

\begin{figure}
\caption{Rotational spectra for ten electrons (upper diagram) and twenty
electrons (lower diagram), when they have formed Wigner molecules at
$B_0= 20 \; {\rm T}$,
as a function of total angular momentum $I$}
\label{fig5}
\end{figure}

\begin{figure}
\caption{Heat capacity $c$ arising from the rotational spectra of
Figure \ref{fig5}. The dashed line shows the twenty electron system,
the solid one the ten electron case.}
\label{fig6}
\end{figure}

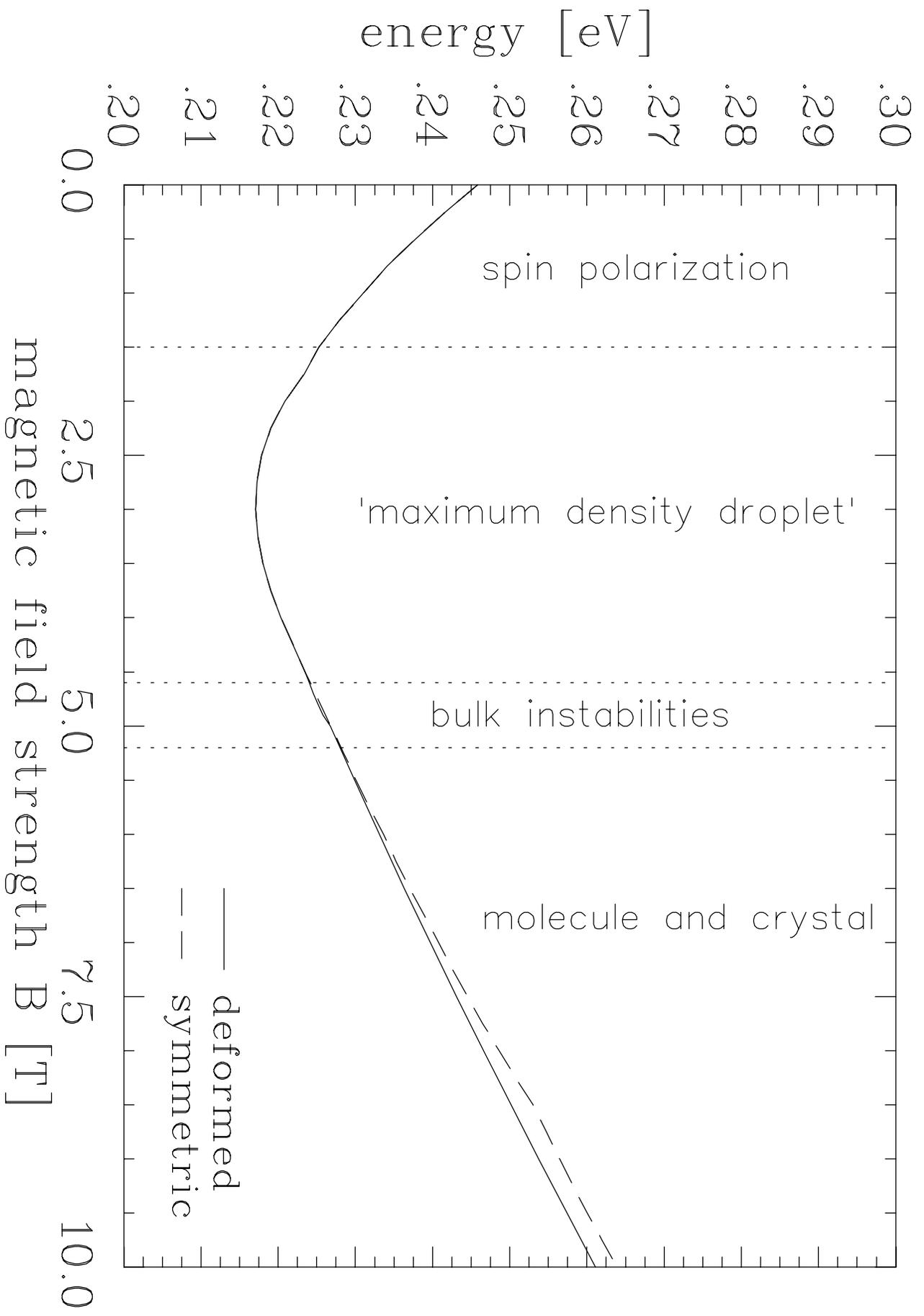
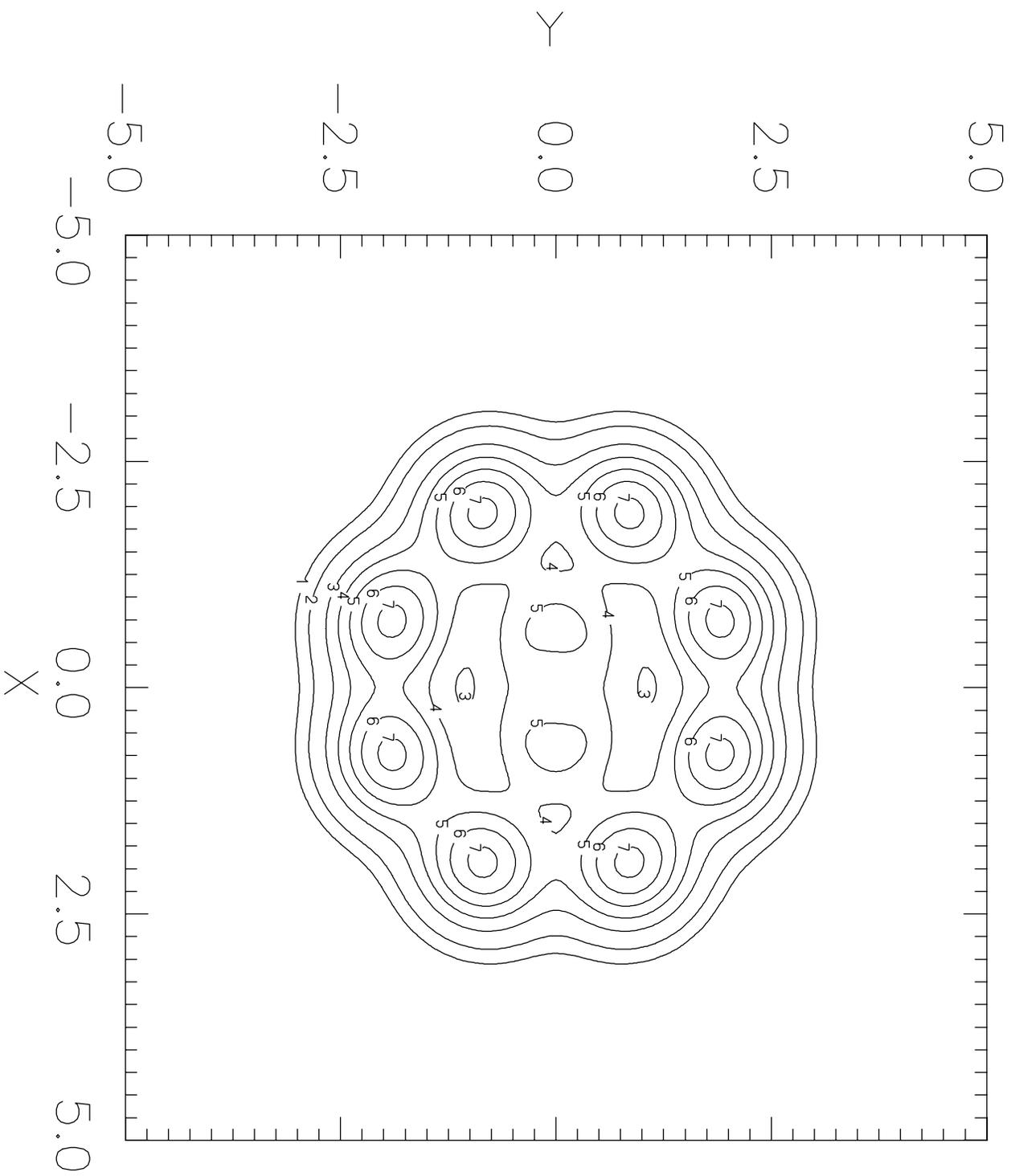
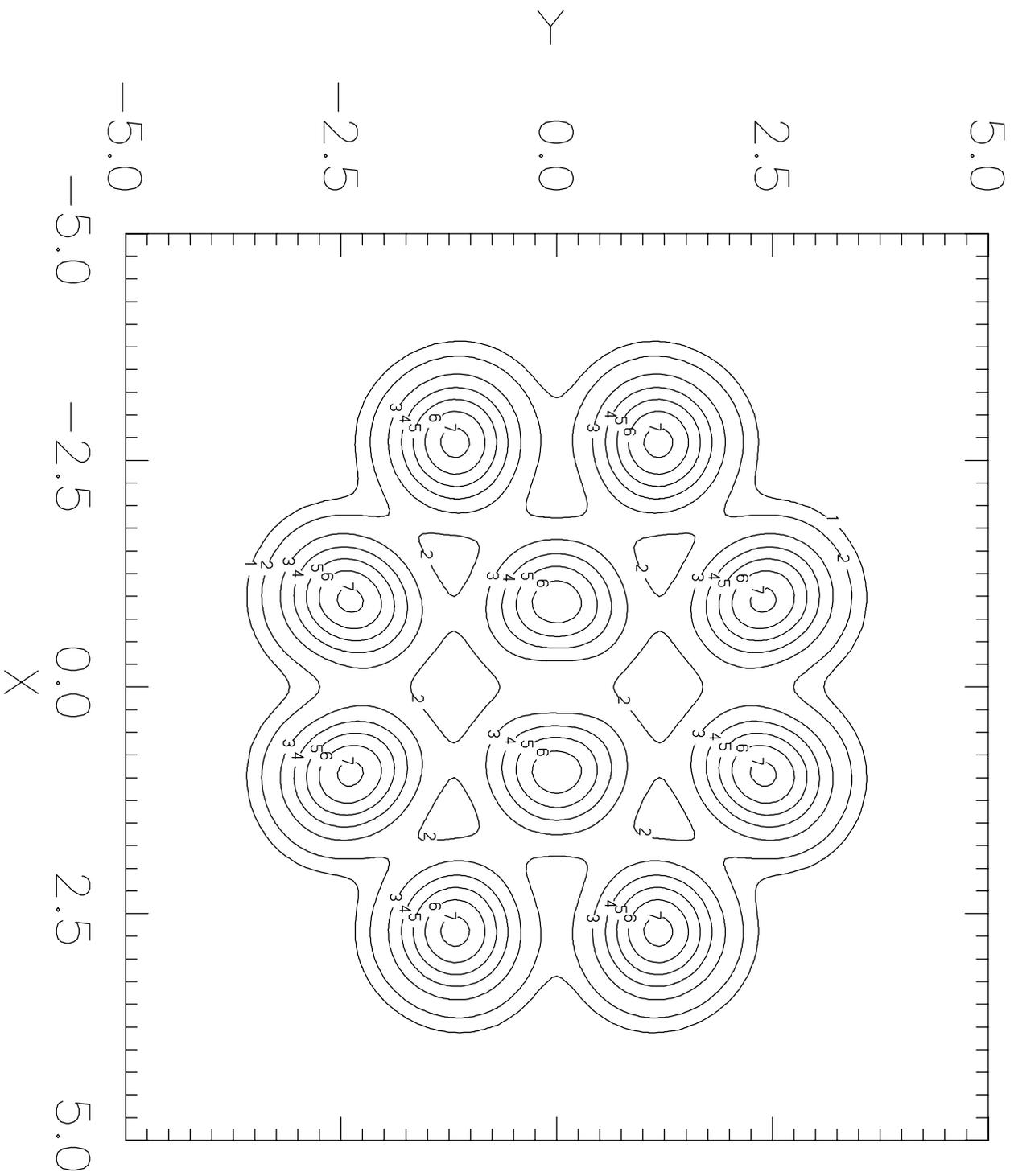
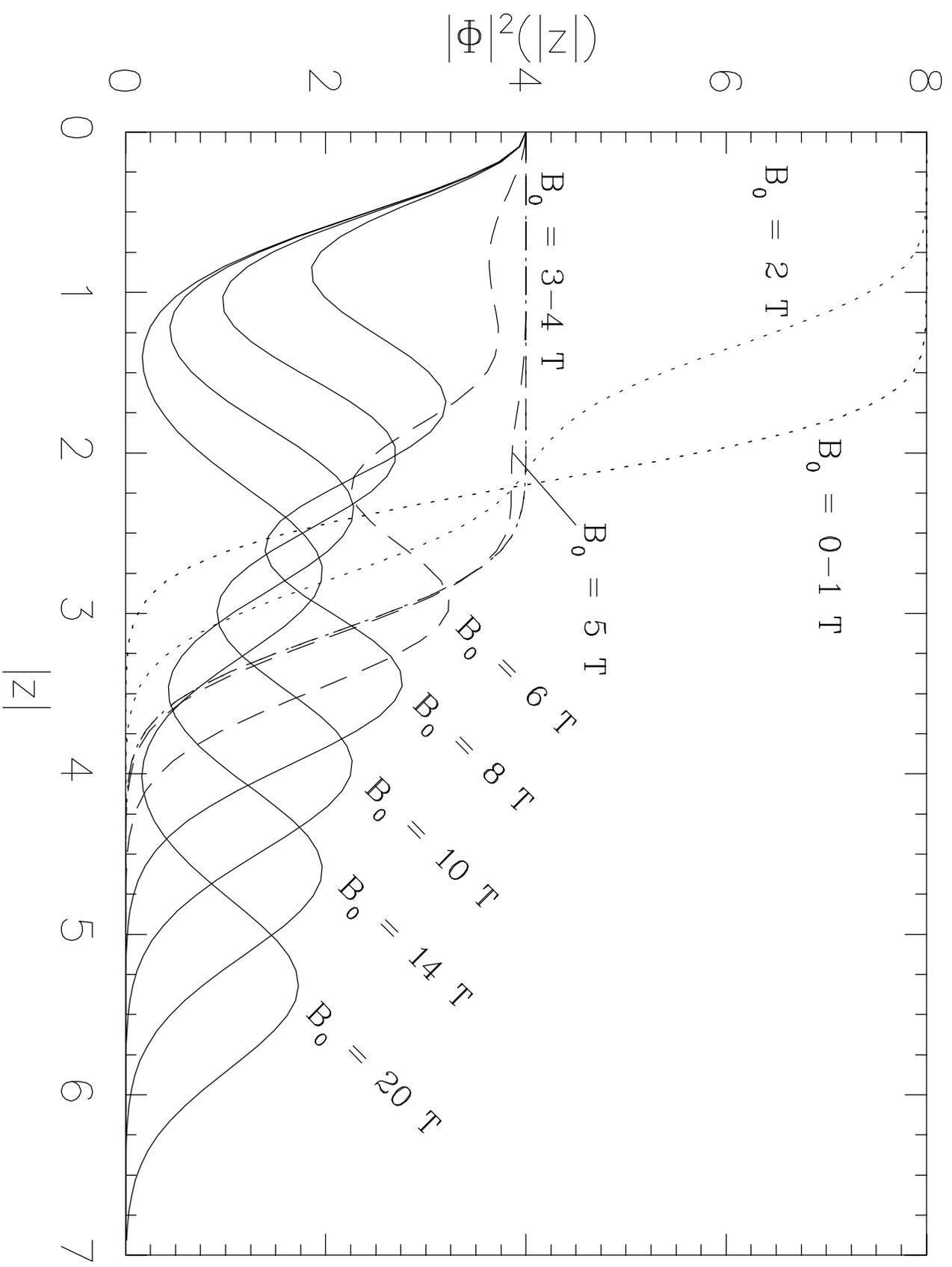
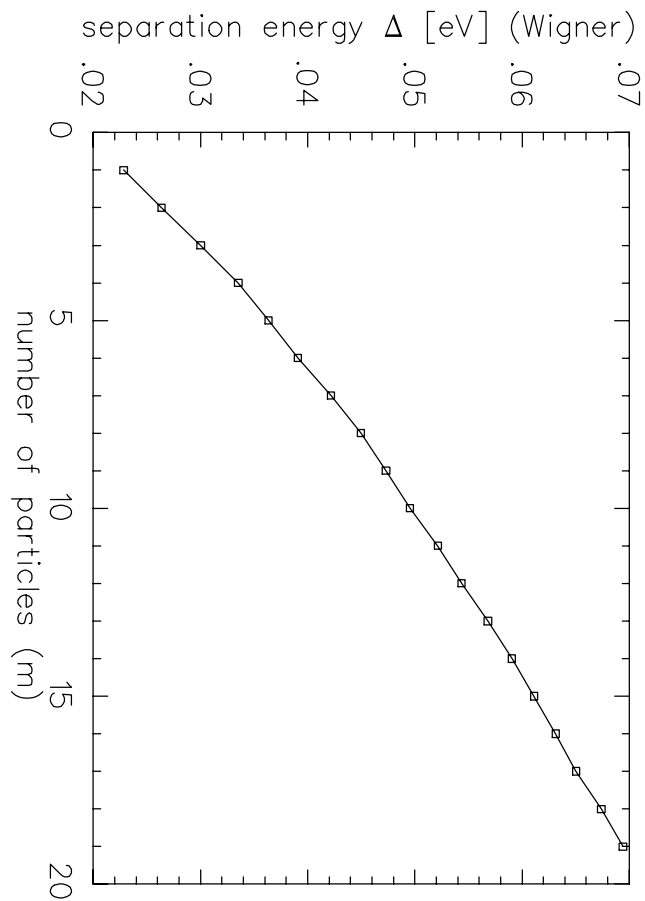
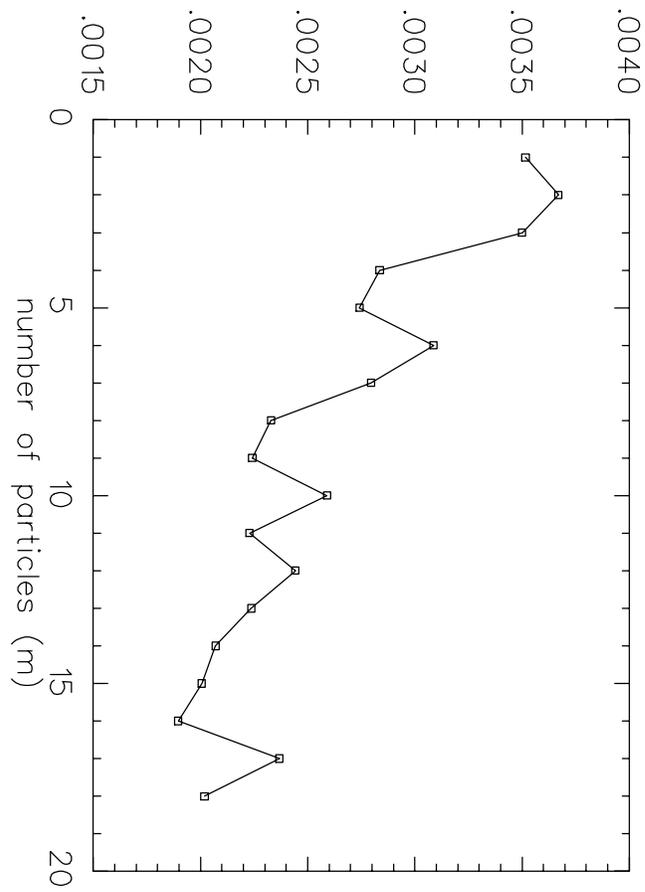
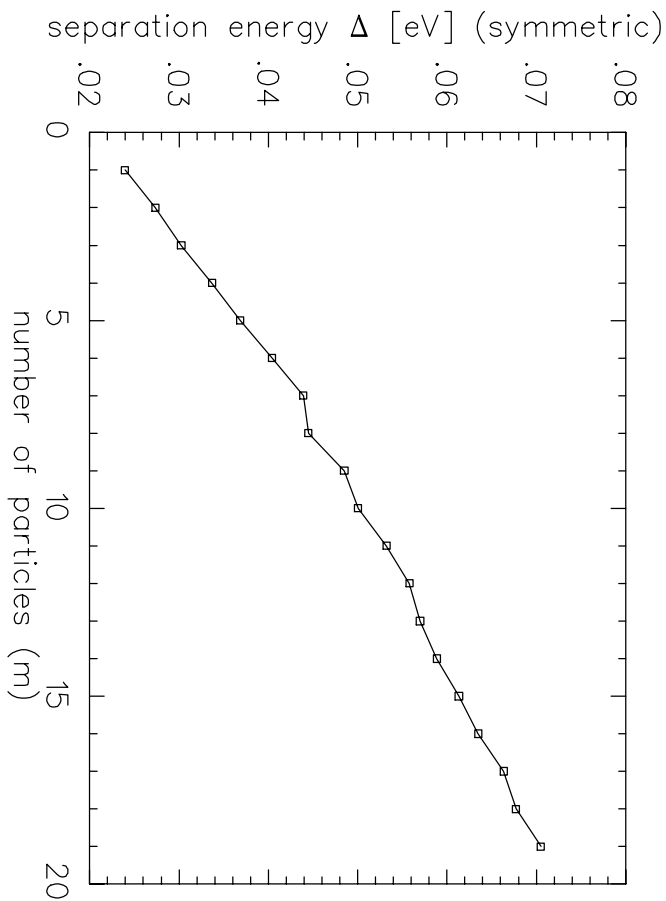
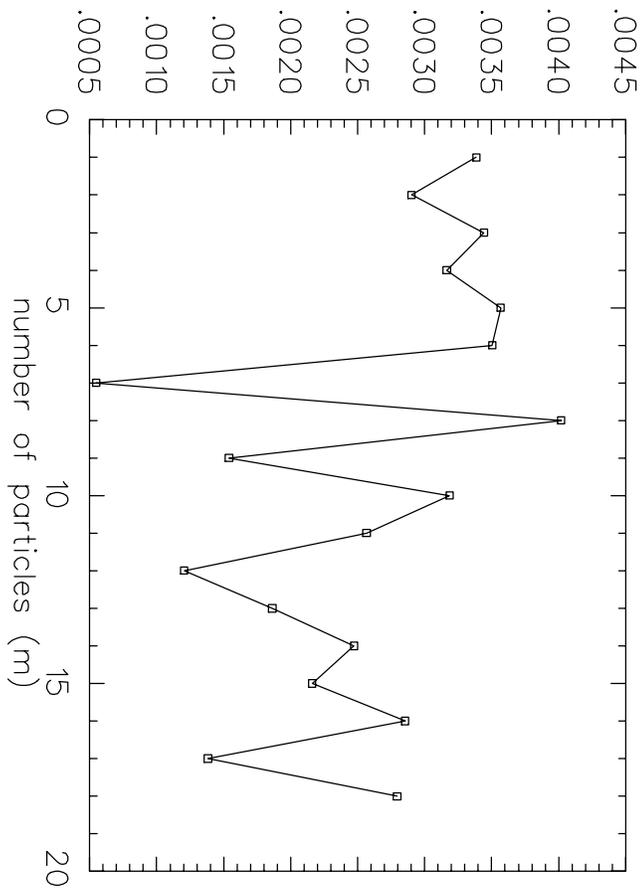
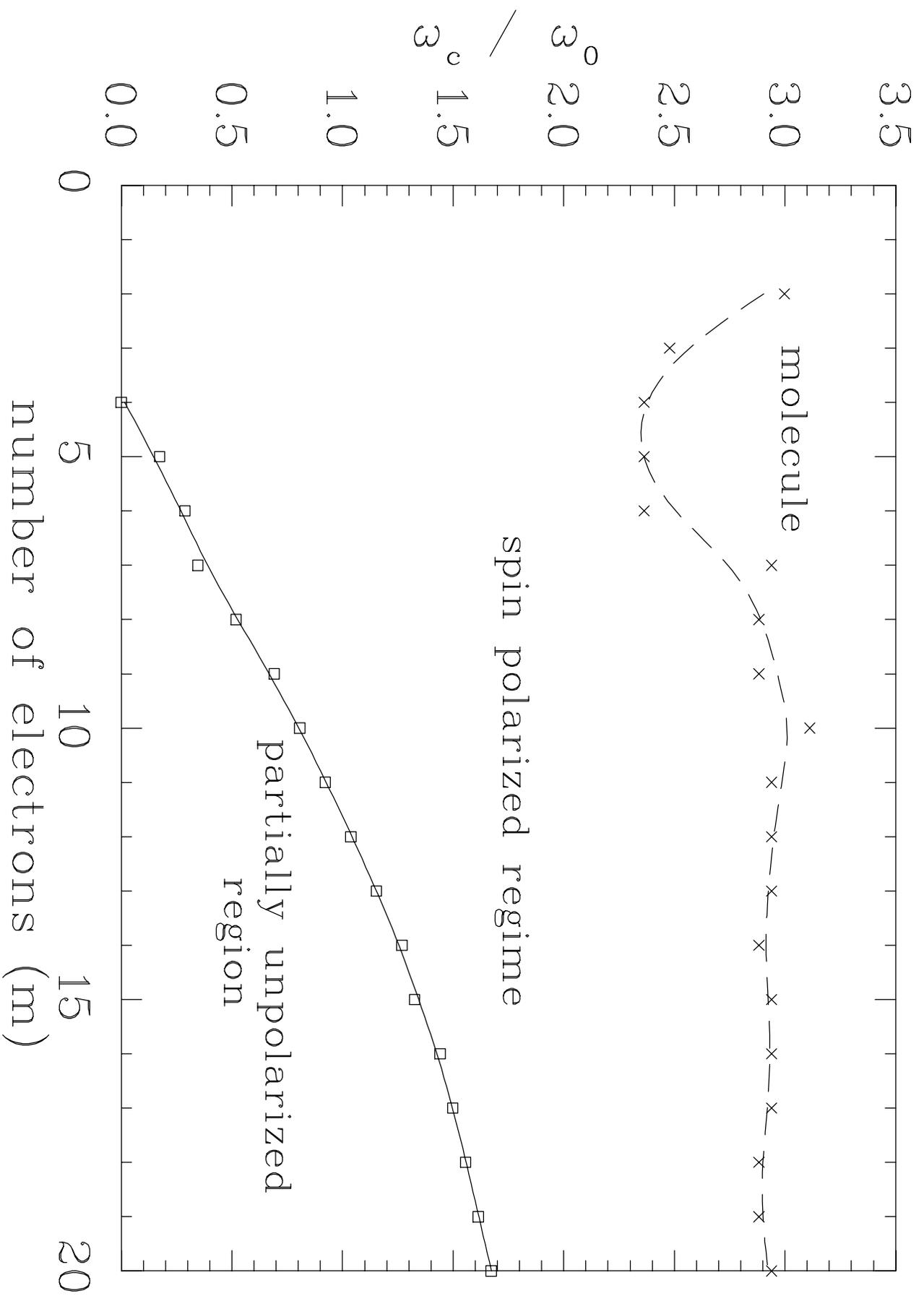
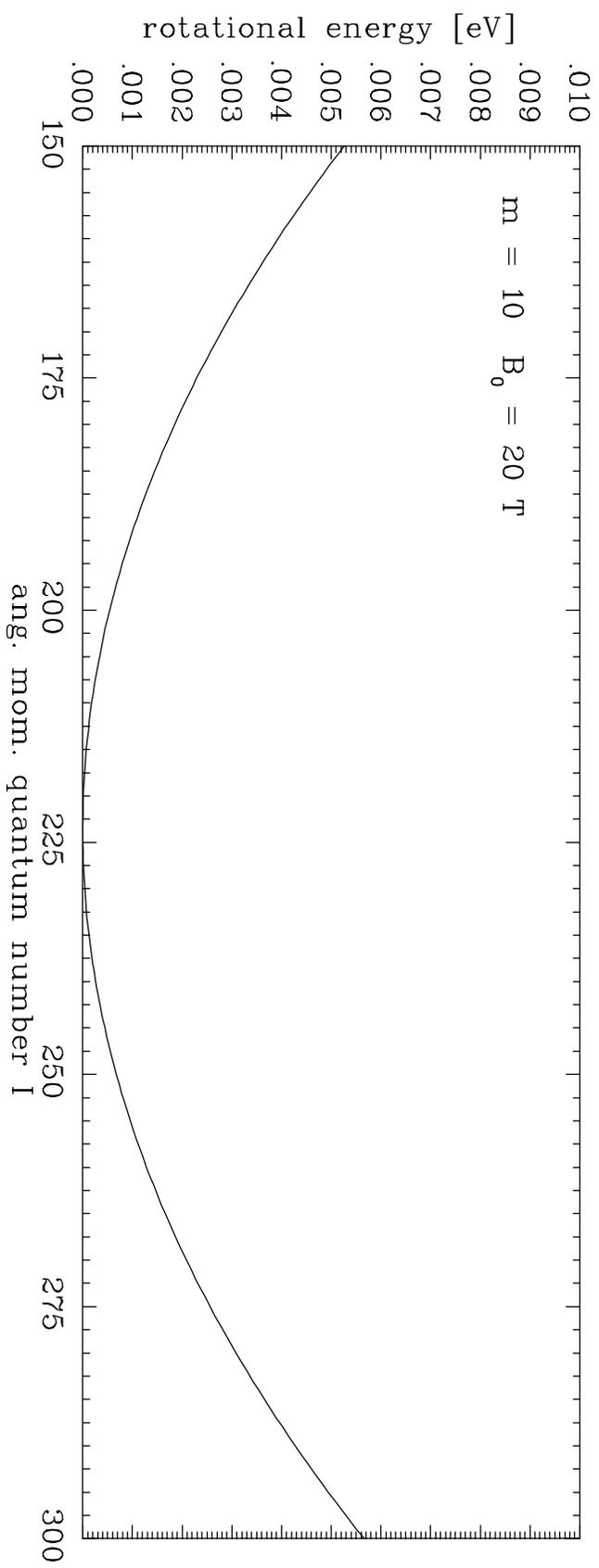
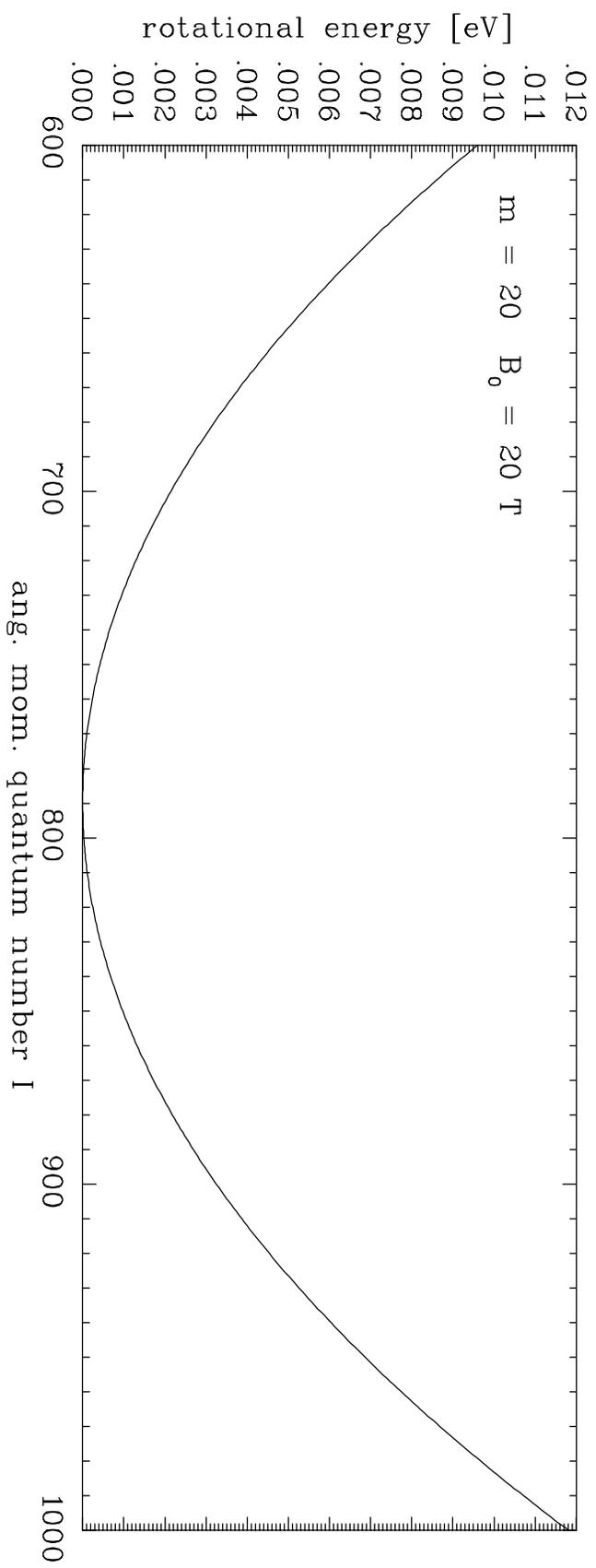
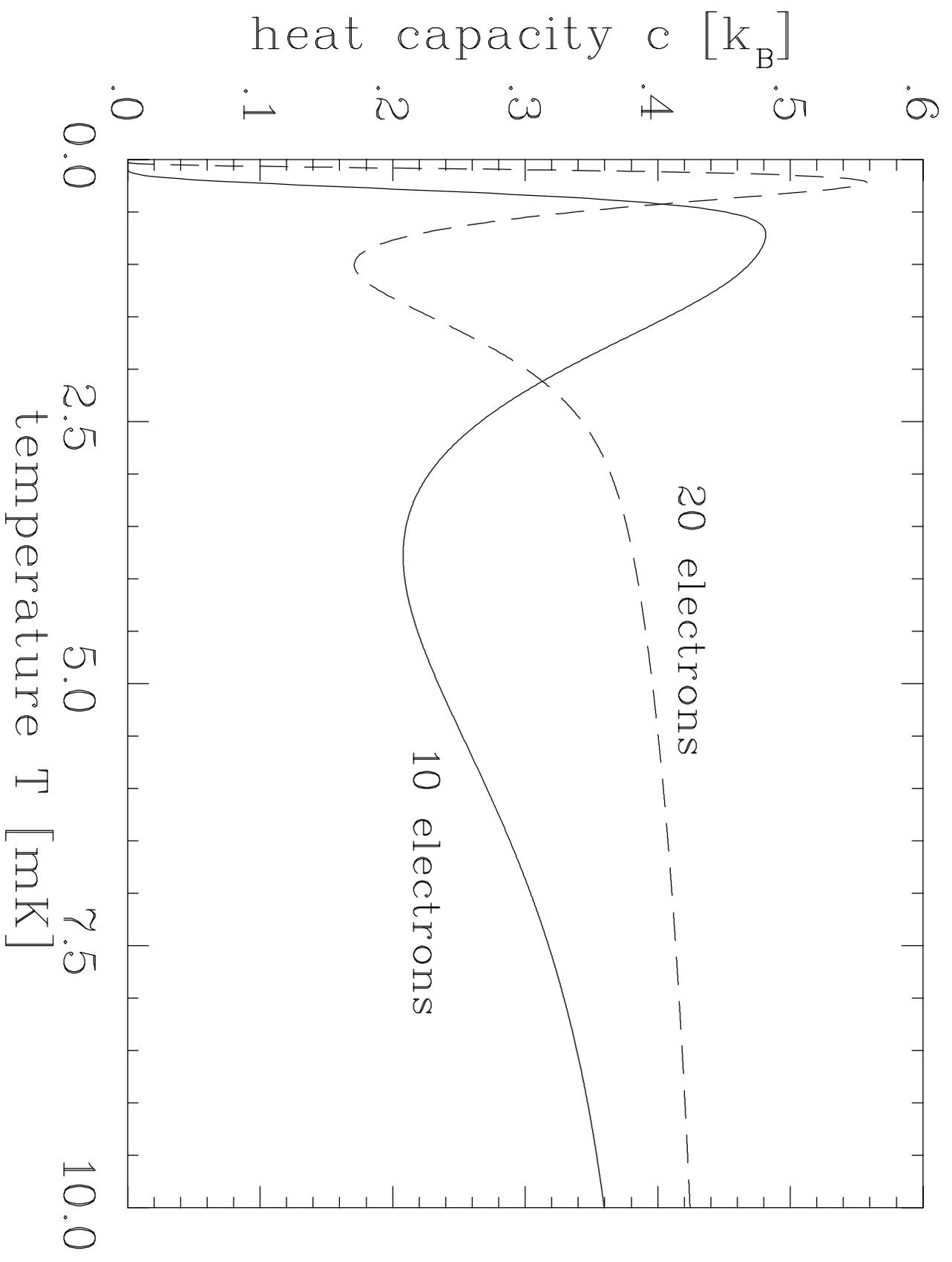
\end{document}